# Strong light enhancement by combining the photonic nanojet and plasmons from the nano-engineered microsphere

*Vlatko Gašparić[1], Mile Ivanda[1]*, Davor Ristić[1], Tamás Váczi[2], Hrvoje Gebavi[1], Miklós Veres[2]*

[1]Laboratory for Molecular Physics and Synthesis of New Materials, Ruđer Bošković Institute, Bijenička cesta 54, 10000 Zagreb, Croatia
[2]Nanostructures and Applied Spectroscopy Research Group, HUN-REN Wigner Research Centre for Physics, Konkoly-Thege Miklós út 29-33, 1121 Budapest, Hungary

*Corresponding author: ivanda@irb.hr

## Abstract

Today's cutting-edge optical spectroscopic exploratory tools, such as Raman or infrared spectroscopy, rely on methods of signal enhancement as a route for their development. These methods are indispensable for substance identification and characterization in almost any scientific, regulatory, or industrial laboratory, therefore new and better methods of enhancement are always sought after. In this paper, the design of a new optical device for enhancement is presented, called nano-engineered microsphere (NMS). This device innovatively combines plasmons, a present flagship enhancement method, with a photonic nanojet, a new and emerging enhancement tool, to provide unprecedented properties in terms of affordability, stability, and performance. By using numerical simulations, a detailed design of the device is presented, and the optimization of device parameters for the strongest enhancement is investigated. The simulations show different influences of the parameters on the enhancement, from low to critical. The most influential parameter was found to be the radius of the nanoelement tip, which, at low values, showed a tremendous increase in the enhancement. The optimized device shows exceptionally promising abilities regarding the enhancement, while the estimated cost of production and use is low. Such properties paired with low price and ease of usage could enable the NMS to become one of the leading methods of enhancement in Raman and infrared spectroscopy with the spatial resolution towards nanometers.

## Introduction

In almost any scientific, regulatory or industrial discipline, substance identification and characterization cannot be imagined without spectroscopic tools such as Raman, infrared or fluorescence spectroscopy. To increase their capabilities, many methods of enhancement have been developed and are constantly being improved. Plasmonic amplification often plays a key role in today's leading methods of enhancement in each spectroscopic tool. Examples are tip-enhanced Raman spectroscopy (TERS)[1,2], surface-enhanced Raman spectroscopy (SERS)[3], surface-enhanced infrared absorption (SEIRA)[4] or plasmon-enhanced fluorescence spectroscopy (PEFS)[5]. Plasmons are the collective oscillations of the conduction electrons in metals, which are excited in suitable conditions when the metal is illuminated with light. They enable tremendously high signal enhancement factors, for example, up to $10^8$ in the case of Raman spectroscopy. For SERS and SEIRA (surface methods), carefully prepared and specific substrates with metallic nanoparticles are needed. The drawbacks of these methods are low reproducibility and complicated fabrication of substrates, which need to be adjusted for the specific analyte to be characterized and for the laser wavelength used. In TERS, plasmons are excited at the metallic tip, which is positioned under the laser beam at the location on the sample where the measurement is performed. Because of that, TERS has better reproducibility compared to SERS, and, due to the nanometer-sized end of the tip, also achieves extremely high resolution in Raman mapping. However, disadvantages of TERS are the high price, complicated design and demanding usage. TERS requires an additional atomic force microscopy (AFM) or scanning probe microscopy (SPM) system integrated with the Raman spectrometer. Often the TERS-enabled systems a have design where the microscope objective is positioned at an angle and the sample is illuminated from the side. This leads to additional problems with sample excitation and signal collection.

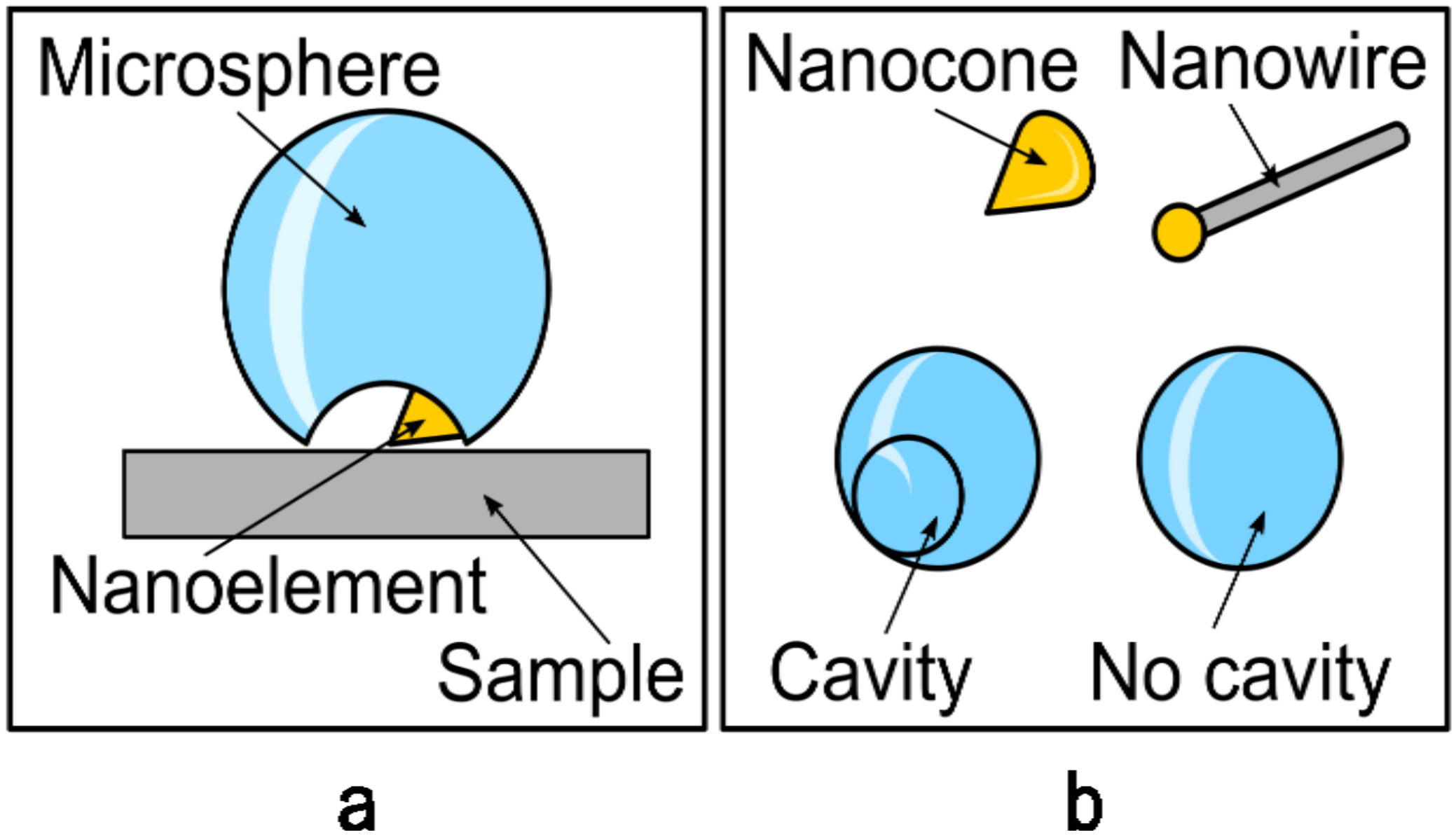

Figure 1. (a) Schematic illustration of the NMS on the sample. The NMS consists of a dielectric microsphere and a nanoelement built in the shadow side of the microsphere. (b) Some examples of nanoelements (nanocone and nanowire) and microspheres (with or without cavity).

One of the newest and most promising enhancement methods in micro-spectroscopy is based on the creation of a powerful and narrow beam of light called photonic nanojet (PNJ). It is generated under suitable conditions at the shadow side of an illuminated microlens (most often a microsphere)[6]. When such a microsphere is positioned and illuminated at a desired location at the sample, the enhancement occurs due to the strong and localized electric field of the PNJ. PNJ has already shown great results in Raman enhancement, as a standalone method[7–11] and also in combination with SERS[12,13]. The combined enhancement by PNJ and SERS proves that a synergistic effect of the PNJ and plasmons is feasible and effective. However, as long as the plasmons are excited at the SERS substrates, the problems, such as low reproducibility and complex preparation of substrates, are still present. The next step would be a combination of the PNJ with TERS, which is so far not reported. Moreover, PNJ has also been successfully used for fluorescence enhancement[14]. When the microsphere position is controlled, the PNJ can also enable better mapping resolution[15]. The PNJ has also been used for various other applications, such as nanolithography[16], super-resolution[17–19], photonic hook[20], optical waveguiding[21], optical forces[22], optical tweezers[23], solar cells[24], data storage[25], and laser surgery[26].

Here we present a new optical device called nano-engineered microsphere (NMS), which combines a powerful PNJ and plasmonic effect at the tip of a metallic nanoelement in a microsphere. The NMS enables the combined enhancement of the PNJ and plasmons, which reaps the benefits of both methods of enhancement but avoids their disadvantages. The NMS can also be viewed also as a unique kind of combination method of the PNJ and TERS, which is currently the only solution in this approach.

**The NMS at a glance**

The NMS consists of an optically transparent dielectric microsphere that has an integrated nanoelement, as shown in **Figure 1a**. The design of the NMS is relatively simple, yet its function

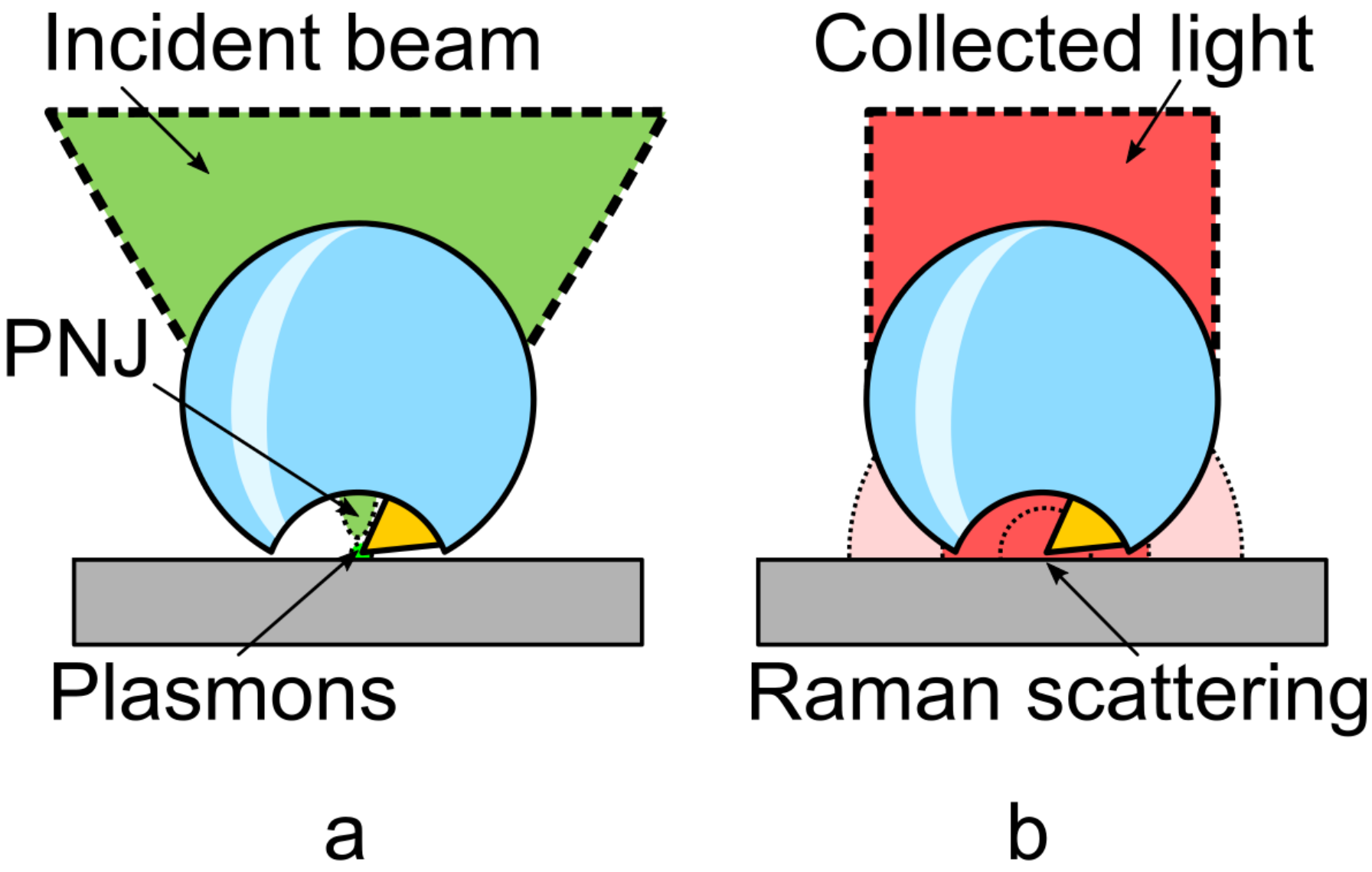

Figure 2. (a) Schematic illustration of incident beam enhancement by NMS. The incident beam is focused by microsphere to generate a PNJ which excites plasmons at the tip of the nanoelement. (b) Schematic illustration of collecting enhancement by NMS. Raman scattering light from the sample is enhanced once more by the plasmons and collected by the microsphere which collimates the light towards the microscope objective.

is fairly complex. This is due to the elegant combination and usage of the components of the NMS. The microsphere performs three major functions in the device. Firstly, it acts as a mechanical holder of the nanoelement. Secondly, when illuminated, it generates a PNJ at its shadow side, which is then projected onto the nanoelement and the sample surface (**Figure 2a**). Thirdly, it functions as a collecting and collimating optical element due to its antenna effect. The microsphere enables the collection of the scattered light from a solid angle subtending almost 180º, and passes the light with very small divergence to the microscope objective (**Figure 2b**).
In the NMS, the plasmonic excitation happens when the PNJ illuminates the tip of the nanoelement. The base of the nanoelement is built in the shadow side of the microsphere, positioned at the side and inclined so that the tip of the nanoelement is placed on the axis of light propagation, at the center of the nanojet. It can be of various shapes, such as a nanocone or a nanowire (**Figure 1b**)
The microsphere can take the form of a complete sphere or alternatively have e.g. a cavity on the shadow side (**Figure 1b**), which provides the benefits of protecting the nanoelement and an improved focusing of the incident light beam

The enhancement by NMS is caused by a synergistic effect of the plasmons and photonic nanojet, which is schematically illustrated in **Figure 2**. The plasmonic excitation generates extremely high levels of enhancement by enhancing both the incident and scattered light from the sample. It has been shown that PNJ enhances the Raman signal in standard micro-Raman scattering configuration by more than one order of magnitude, even when using 10× to 50× microscope objective[10,27]. Moreover, it also reduces the size of the laser spot diameter on the sample by 2-3 times [28,29], which can result in highly increased laser power density. Therefore, the PNJ not only excites the plasmons at the tip of the nanoelement, but also enhances the Raman signal itself. Furthermore, the antenna effect of the microsphere is an additional contribution to the enhancement, which also enables ultra-long working distance confocal collection of scattered light[30].

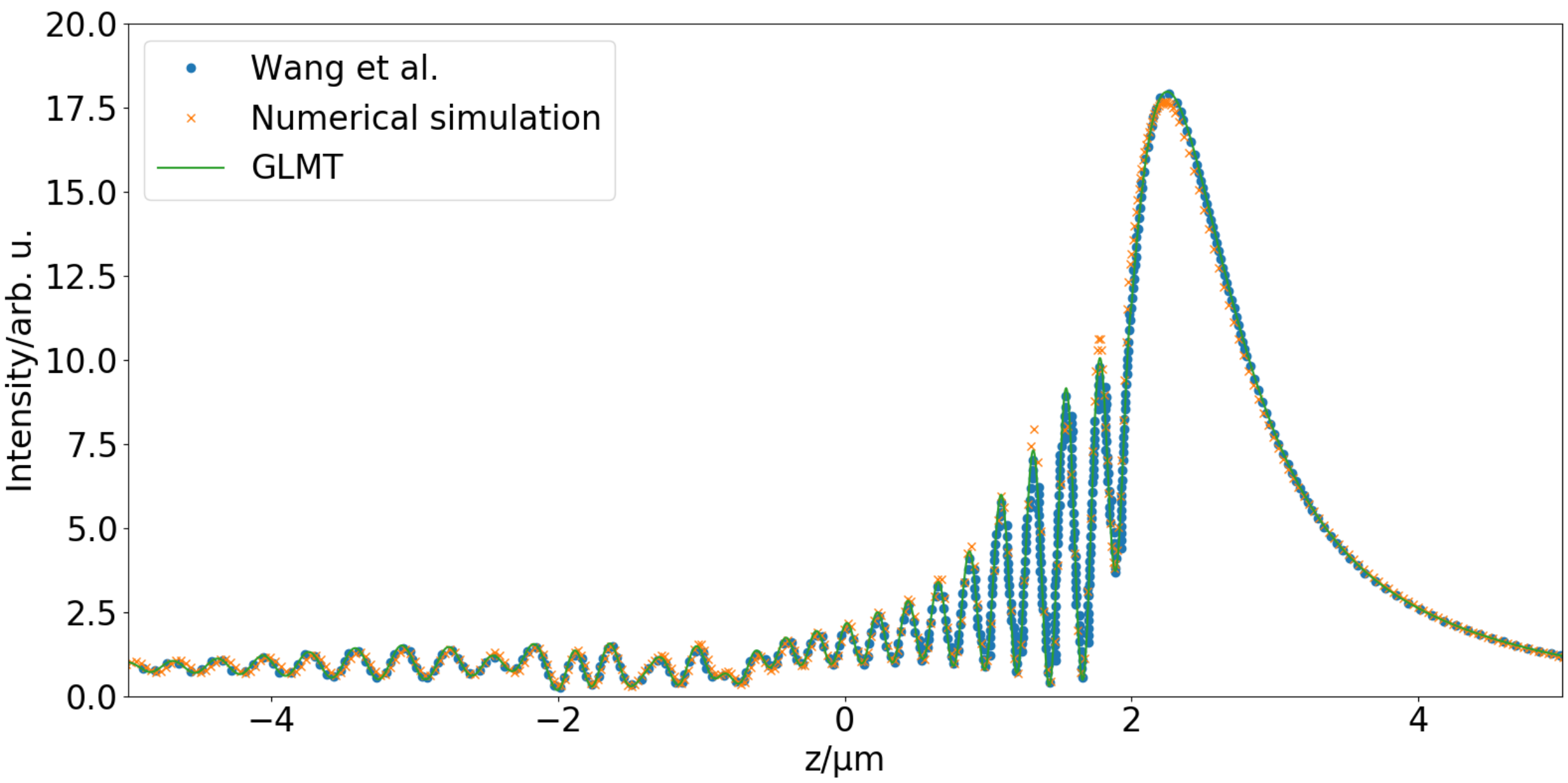

Figure 3. Validation of the numerical model: plot showing the intensity of the electric field along the axis of light propagation through a microsphere of 2 μm radius. The model is compared with our own GLMT calculations and the GLMT calculations from the literature.

**Numerical model**

The 3D finite-difference time-domain (Ansys Lumerical FDTD) simulation model consisted of the NMS in vacuum upon which a Gaussian beam propagating along the vertical z-axis was projected. The microsphere was positioned at the center of the coordinate system. Boundary conditions of the simulation space were set as perfectly matched layers. The simulation temperature was set at 300 K. The simulation time was mainly set at 100 fs with an automatic shutoff triggered if the energy inside the simulation space droped below a factor of $10^{-5}$. In some configurations, longer simulation time was used, as needed. The simulation mesh was automatic, non-uniform, and generated by the software for each configuration. The mesh was fine at the positions where the cavity and the nanoelement were, and coarse further away and outside the microsphere. To achieve even higher simulation resolution, a custom uniform mesh with a 1 nm step was added at the nanoelement tip position, which overrided the automatic mesh.

**Numerical simulations validation**

Before performing the optimization of the NMS parameters by the FDTD numerical simulations, a validation of the numerical model was done in order to ensure the accuracy and validity of the model. The validation was done by comparing the numerical model with our own Generalized Lorenz-Mie theory (GLMT) calculations and with the GLMT calculations from the literature[31]. Also, a series of simulations of the same NMS configuration with different simulation parameters were performed in order to obtain adequate simulations parameters such as simulation time, boundary conditions, cutoff value, mesh size, etc.

The comparison of the simulation and calculations is shown in **Figure 3** where the photonic nanojet emerging from the microsphere of a radius of 2 μm is shown by plotting the intensity of the electric field on the axis of light propagation through the microsphere (z-axis). The results obtained by our own and literature GLMT calculations are completely matched, which is to be expected since GLMT is an analytical method of calculation. This confirms the validity of our GLMT model, which was important because the numerical model was further validated by out GLMT model. The numerical results show very good match with the GLMT method. After fine tuning the parameters of the simulation model, an optimal model was found regarding the mesh size and simulation time and our numerical model is validated to be sufficiently accurate to proceed with the optimization of the NMS parameters.

**NMS parameters variation**

The design and function of NMS can be tailored by changing the parameters of the microsphere and the nanoelement. In order to optimize the design and the performance of the device, variations of the parameters of the device were performed. The parameters were separated into two groups: firstly, the parameters of the incident beam (wavelength $\lambda$, beam waist radius $w_0$ and beam waist position $z_0$) and the microsphere (radius $R$ and refractive index $n_s$); and secondly the parameters of the cavity (radius $r_c$ and distance from the microsphere center $z_c$) and the nanoelement (angle relative to the vertical axis $\alpha$, inner angle $\theta$, length $h$, complex refractive index $n_n$, and radius of the curvature of the tip $r_t$).

The schematic of all the parameters can be seen in **Figure 4**. Since there are many parameters to optimize (5 parameters in the first group and 7 parameters in the second group) and numerical simulations are both time and memory demanding tasks, a careful strategy was developed in order to cover as many configurations as possible, while keeping the number of configurations tolerably low. The first step in the optimization was to define a "standard configuration" – a configuration whose parameter values are realistic, usual for a future experimental implementation, which were, through previous research, shown to be reliable or those which are in the middle of the range of possible values. The standard configuration hence was defined with these values: $\lambda = 532$ nm, $w_0 = 0.7$ μm, $z_0 = 10$ μm, $R = 2.5$ μm, $n_s \rightarrow SiO_2$, $r_c = 125$ nm, $z_c = 2500$ nm, $\alpha = 60°$, $\theta = 45°$, $h = 250$ nm, $n_n \rightarrow$ Ag and $r_t = 10$ nm.

In the second step, from the standard configuration, the parameter variation ranges were defined, which are shown in **Table 1**. The parameter variation ranges were chosen to cover a broad range of possible future experimental implementations, with steps fine enough to provide trends and dependencies in graphs but coarse enough to be manageable within time and memory constraints. The wavelength variation covered all the usual wavelengths used in TERS experiments, from near UV to near IR. The incident beam size values corresponded to beams which would emerge from the usual microscope objectives from 100× to 10×, while the size and the material of the microsphere were chosen to be appropriate for the photonic nanojet generation. The values of the geometrical parameters of the cavity and the nanoelement were limited by the geometry of the device, while the material of the nanoelement covered possible candidates for a high plasmonic excitation by the visible light.

In the third step, the variations were performed from the standard configuration by varying one parameter at a time while other parameters were fixed.

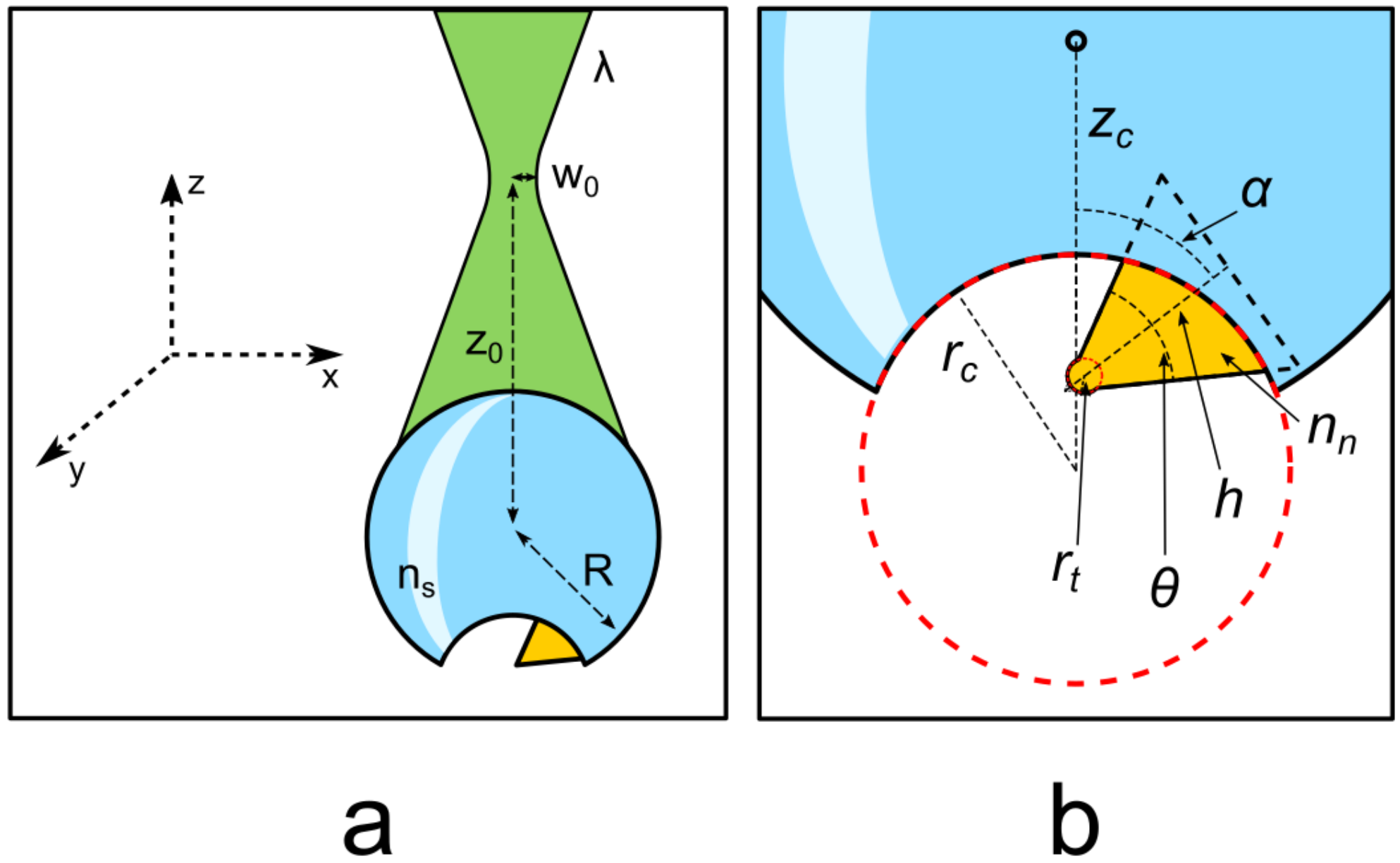

Figure 4. Parameters of the NMS divided into two groups: (a) Parameters of the incident beam (wavelength $\lambda$, beam waist radius $w_0$ and beam waist position $z_0$) and the microsphere (radius $r$ and refractive index $n_s$); (b) Parameters of the cavity (radius $r_c$ and distance from the microsphere center $z_c$) and the nanoelement (angle relative to the vertical axis $\alpha$, inner angle $\theta$, length $h$, complex refractive index $n_n$, and radius of the curvature of the tip $r_t$).

The parameter $z_0$ is more special than other parameters since experimentally it would be determined by positioning the microscope objective height relative to the microsphere and thus the optimum in simulations was determined in different ranges for each configuration when the first group of parameters were optimized.

Table 1. The values of the parameters used for the optimization of the NMS device. The values of the parameters of the standard configuration are underlined.

| $\lambda$/nm | $w_0$/µm | $R$/µm | $n_s$ | $r_c$/nm | $z_c$/nm | $\alpha$/ ° | $\theta$ / ° | $h$/nm | $n_n$ | $r_t$/nm |
|---|---|---|---|---|---|---|---|---|---|---|
| 351 | 0.5 | 0.5 | $H_2O$ | 0 | 2376 | 22.5 | 10 | 135 | Ag | 1 |
| 488 | 0.7 | 1 | $SiO_2$ | 10 | 2380 | 30 | 20 | 180 | Al | 1.2 |
| 532 | 1 | 2.5 | PS | 25 | 2390 | 40 | 30 | 250 | Au | 1.5 |
| 633 | 2 | 5 | BTG | 50 | 2410 | 50 | 45 | 350 | Cu | 2 |
| 785 | 5 | | diamond | 125 | 2450 | 60 | 60 | 450 | Fe | 3 |
| | | | rutile | 250 | 2500 | 65.5 | 75 | 600 | Pt | 5 |
| | | | | 500 | 2550 | | 90 | 2500 | W | 10 |
| | | | | 1250 | 2580 | | | | | 20 |
| | | | | 2500 | 2600 | | | | | 30 |
| | | | | | 2610 | | | | | 50 |

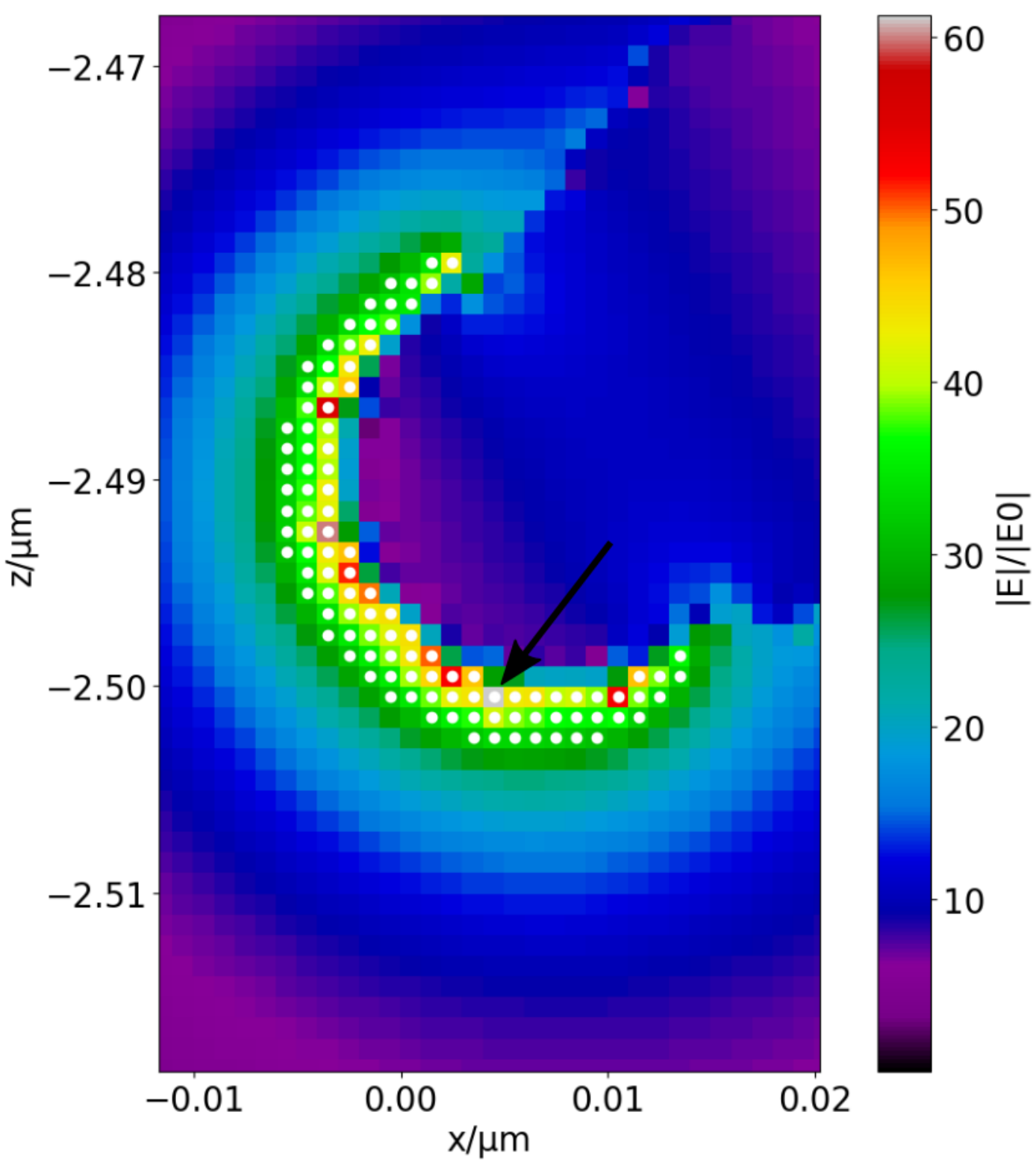

Figure 5. Simulated enhancement distribution around the nanoelement tip. The maximum enhancement point is indicated by an arrow and the top 50% enhancement points are indicated by white dots.

To reduce the optimization time of $z_0$, firstly the range of values for each configuration was reduced by GLMT calculations. Then, out of this reduced range of values, the optimal $z_0$ was determined for each configuration in the first group. In total, more than 60 configurations of the NMS were simulated without counting the $z_0$ parameter variation, or over 120 configurations including the $z_0$ parameter optimization.

**Calculation of the enhancement**

After the completion of a simulation, the electric field enhancement was plotted inside and around the NMS, i.e. at each point of the simulation mesh. The electric field enhancement was gained by normalizing the electric field to the maximum electric field value of the incident Gaussian beam. The enhancement of the NMS device was represented by two values: by the maximum electric field enhancement (single point of a mesh), and the average of the top 50% enhancement values of the enhancement distribution around the nanoelement tip. The enhancement distribution around the nanoelement is shown in **Figure 5** where the maximum enhancement point is indicated by an arrow and the top 50% enhancement points are indicated by white dots. These two values provided a representation of the performance of the NMS regarding the enhancement.

**Results**

Numerical simulation result of the standard configuration of the NMS device is shown in **Figure 6**. In **Figure 6a** the light scattering and enhancement is visible inside and around the device. The incident light undergoes Mie scattering on the microsphere and the PNJ is generated at the shadow side (bottom) of the microsphere, where the nanoelement is located. The PNJ excites plasmons at the nanoelement tip, producing a highly enhanced electric field peaking at about 60× enhancement at the very surface of the tip (**Figure 6b**). The enhancement gradually fades with distance from the nanoelement surface, where at 20 nm distance from the tip surface it still retains a formidable enhancement of around 10×. It is noteworthy that the enhancement is distributed almost uniformly the whole exposed tip (top, sides and bottom) and not just the sides like in usual single or double nanoparticle plasmon simulations. This is important for the application of the device because it will enable higher enhancement and more coverage of the enhancement distribution relative to the analyte molecules at the surface of the sample.

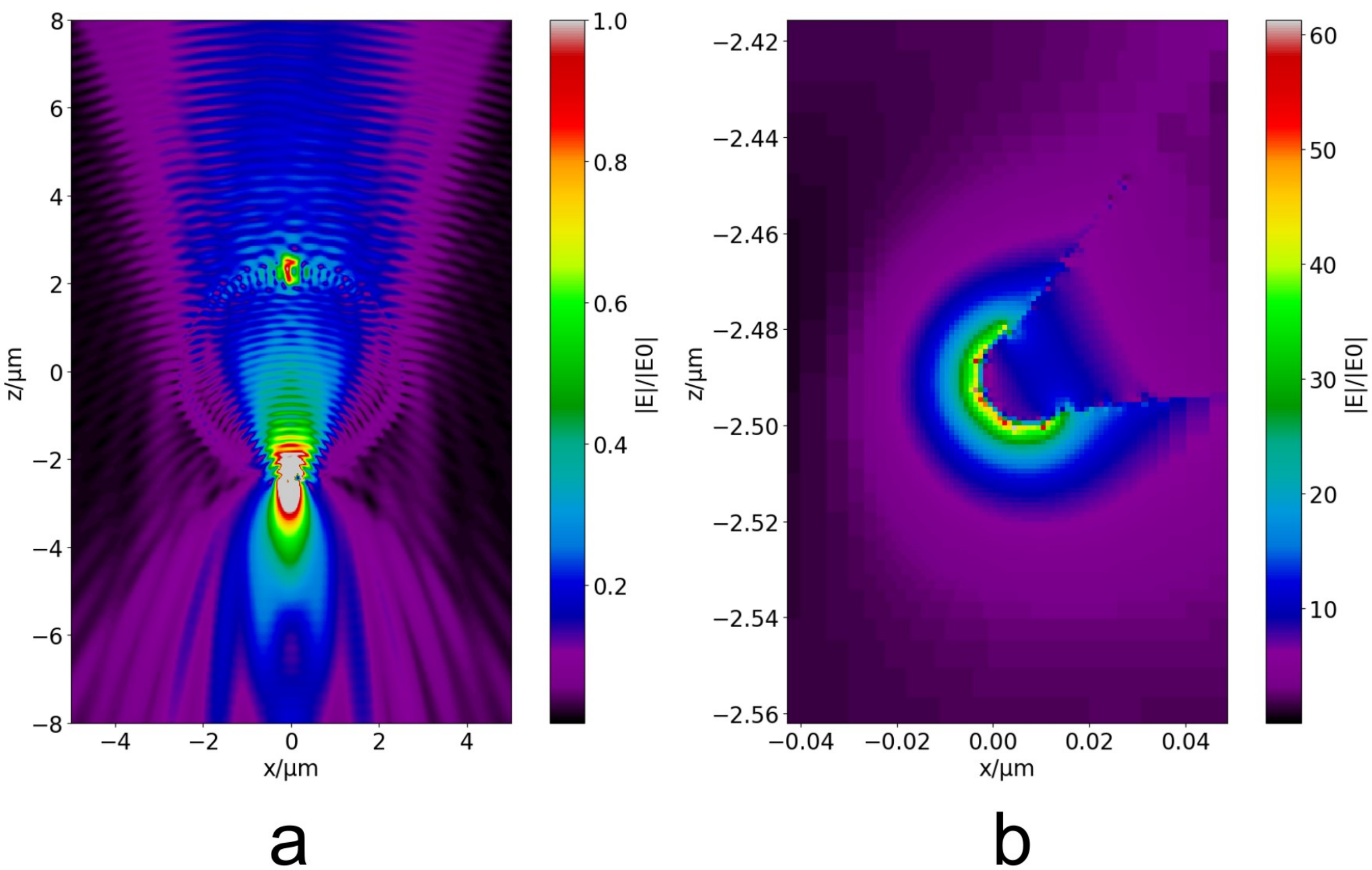

Figure 6. Numerical simulation of the standard configuration of the NMS device. The light propagation is in −z direction. (a) x-z plane image of the electrical field enhancement inside and around the NMS device. The colorbar is intentionally clipped at value 1.0 in order to show the details of the image outside of the hotspot area of the nanoelement tip. (b) x-z plane image of the electrical field enhancement inside and around the nanoelement tip. The highest enhancement of around 60× is present at the tip surface and it gradually fades with distance.

The variation of the first group of parameters is shown in **Figure 7**. The dependence of the enhancement on parameters is shown in two ways: by the maximum electric field enhancement (single point) (blue) and the average of the top 50% enhancement values of the enhancement region around the nanoelement tip (orange). The 50% threshold was chosen as a good tradeoff between covering a large enough plasmonic region and filtering high enough intensity points. Firstly, it is visible that the two representations of the enhancement are in good agreement. They are following the same dependencies, while there is only a relative difference in values. The enhancement shows different behaviors depending on the varied parameter. In the case of microsphere refractive index $n_s$ and incident beam waist radius $w_0$, the enhancement is the highest for their low values and significantly decreases when those parameters increase. On the other hand, the wavelength dependence has a maximum around 650 nm, while it is strongly quenched for short wavelengths such as 400 nm. Regarding the microsphere size, the enhancement remains relatively constant as microsphere radius changes its value with some preference for only small-sized microspheres. Based on these variations, the optimal configuration would consist of a silica microsphere with radius of around 1μm, while the incident beam would need to have the smallest possible waist and a wavelength of around 650 nm.

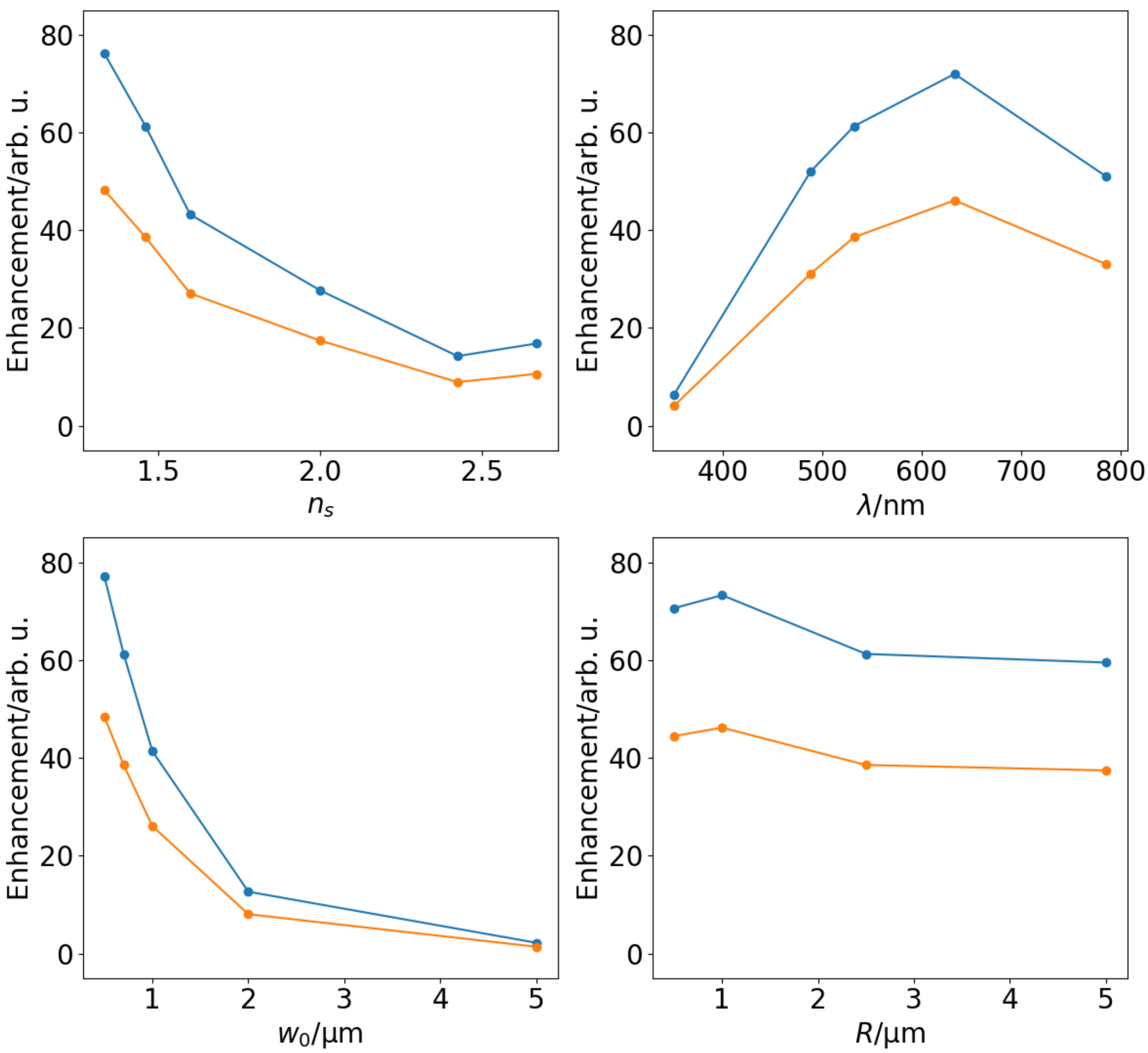

Figure 7. Dependence of the enhancement on the variation of the first group of parameters: incident beam wavelength $\lambda$ and radius $w_0$; microsphere radius $R$ and refractive index $n_s$. The dependence of the enhancement on parameters is shown in two ways: by the maximum electric field enhancement (single point) (blue) and the average of the top 50% enhancement values of the enhancement distribution around the nanoelement tip (orange).

The variation of the second group of parameters is shown in **Figure 8**. Once again, the two representations of the enhancement are in good agreement. Here also the dependencies of the enhancement on the parameters are diverse.

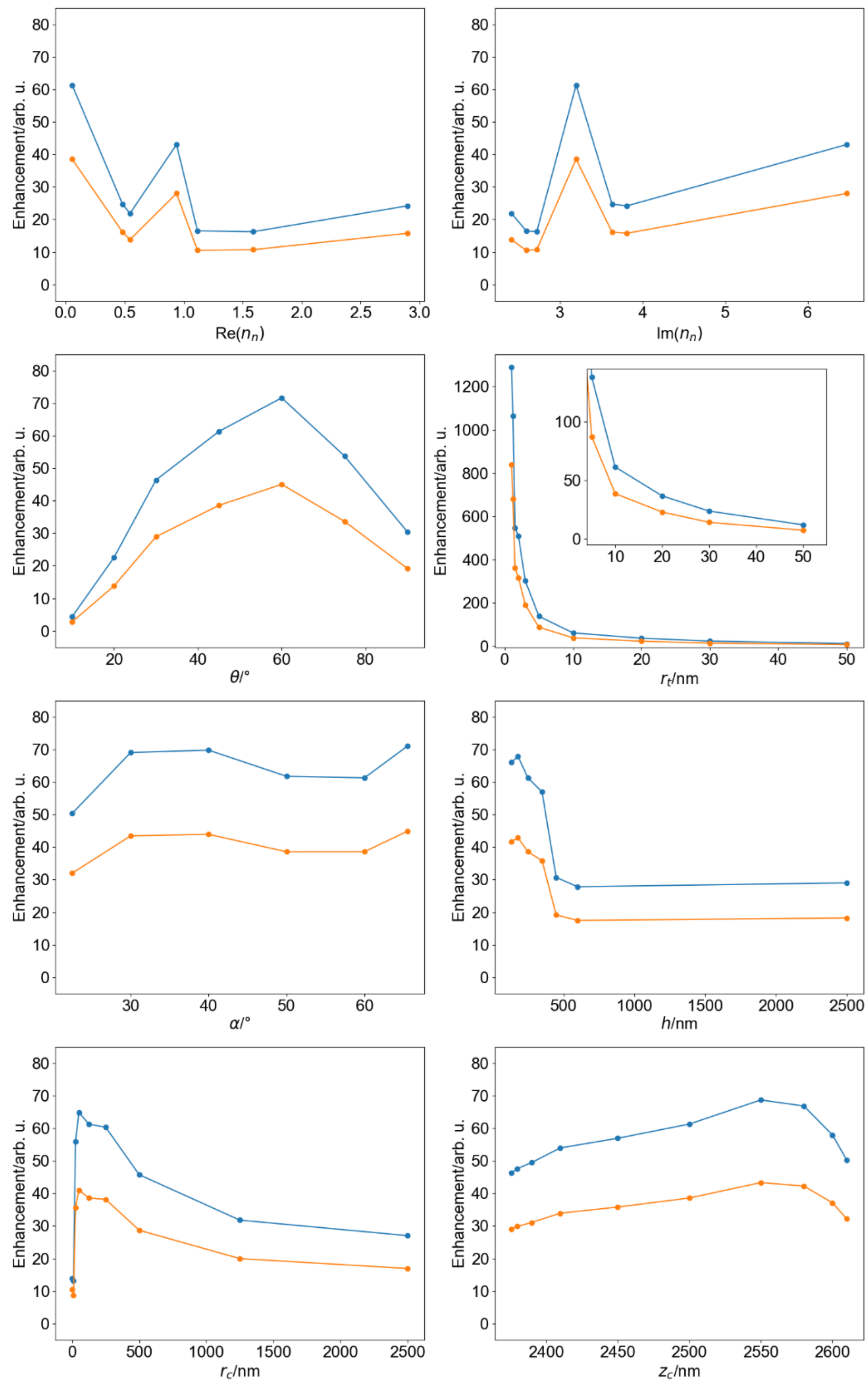

Figure 8. Dependence of the enhancement on the variation of the second group of parameters: the cavity parameters (radius $r_c$ and distance from the microsphere center $z_c$) and the nanoelement parameters (angle relative to the vertical axis $\alpha$, inner angle $\theta$, length $h$, refractive index $n_n$ (real and imaginary part), and radius of the curvature of the tip $r_t$). The dependence of the enhancement on parameters is shown in two ways: by the maximum electric field enhancement (single point) (blue) and the average of the top 50% enhancement values of the enhancement distribution around the nanoelement tip (orange).

The angle of the nanoelement $\alpha$ and the cavity center position $z_c$ have the smallest influence on the enhancement, where $\alpha$ has a small maximum and minimum, and $z_c$ has a small maximum. Therefore, whether the nanoelement tip points in the direction being more parallel ($\alpha = 22.5°$), or more perpendicular ($\alpha = 65.5°$) to the light propagation direction (-z), the enhancement will not change much. The optimum would be between 30º and 40º, and above 60º. Similarly, whether the cavity center is positioned heavily inside the microsphere so that there is barely an opening on the edge of the microsphere ($z_c = 2375$ nm), or the cavity center is at the edge of the microsphere ($z_c = 2500$ nm), or the cavity center is heavily outside the microsphere ($z_c = 2610$ nm), the enhancement will stay relatively similar. The optimum is reached around $z_c = 2550$ nm when the cavity center is slightly outside the microsphere.

On the other hand, the nanocone's inner angle $\theta$ shows a great influence on the enhancement value: very low angle values of 10º or 20º, resulting in a wire-shaped nanoelement, heavily quench the enhancement. This also applies when $\theta$ approaches 90º. The highest enhancement is acquired when $\theta = 60°$, which is a moderate value of this parameter.

The cavity radius $r_c$ and the length of the nanoelement $h$ have similar behaviors, where a narrow range of values has been identified in order to have a substantial enhancement. Otherwise, the enhancement would be very low. For $r_c$, the optimal range is from 25 nm to 250 nm, i.e. the cavity needs to exist and be large enough to accommodate the tip, but below the size which is in the same order of magnitude as the microsphere (in this case up to around 500 nm). For $h$, the optimal range is from 135 nm to 350 nm, which is, similarly to the case of the $r_c$, in the order of magnitude smaller than the microsphere size.

The nanoelement refractive index $n_n$, its both real and imaginary part, do not have a clear trend regarding the influence on the enhancement. The best performing materials were silver and aluminum, while other materials such as platinum, gold, or iron, regardless of the value of the refractive index, provided low enhancement in comparison. However, this parameter is expected to be heavily reliant on the incident wavelength value and the shape and size of the nanoelement tip, since the plasmon excitation on a certain material is primarily influenced by the excitation wavelength and the shape and size of the nanoparticle. Therefore, the influence of the material of the nanoelement should be interpreted in conjunction with light wavelength and tip geometry.

The radius of the nanoelement tip $r_t$ has, by far, the biggest influence on the enhancement, and the trend is following an exponential-like curve. The smaller the $r_t$, the stronger the enhancement, while by increasing $r_t$, the enhancement drops exponentially. Even if it is experimentally almost impossible to realize a tip with a 1 nm radius, it is interesting to see that it would theoretically provide a maximum enhancement of an order of magnitude of 1000×. The tip of a 3-5 nm radius, which would be more experimentally feasible to achieve, would provide a few hundred times enhancement. On the other hand, for $r_t = 50$ nm, the enhancement would drop to around 10×.

**Discussion**

The results showed that the investigated parameters affect the enhancement by the NMS to a different extent. Next, to examine more easily the influence of each parameter, two groups were each subdivided into two, so there are: parameters of the incident beam, the microsphere, the cavity, and the nanoelement.

1. The incident beam strongly affects the enhancement. The effect of the wavelength on the enhancement by the NMS is vastly different than in the case of the PNJ generation alone. The latter can be seen in our previous research about the influence of the parameters on the PNJ,[32] where the dependence of the PNJ intensity on $\lambda$ is a monotonous and significant decrease as $\lambda$ increases, for any incident beam position $z_0$, including the optimal ones. The contrast with **Figure 7** indicates that the dependence of NMS enhancement on $\lambda$ is driven by different factors that overshadow the PNJ contribution. The primary factor is the plasmonic excitation at the nanoelement tip. Here the dependence is linked with the chosen material of the nanoelement, which in this example was silver, and with the shape and size of the nanoelement tip. The configuration of these parameters has led to the dependency we see in the results. Furthermore, this means the dependency on $\lambda$ could be tailored by choosing a different configuration, and that the investigation of the effect of the wavelength on NMS enhancement should be performed with the linked parameters in mind.
The dependence on the incident beam waist radius $w_0$ is similar to the one published earlier.[32] The reason for this behavior lies in the geometry of the beam and the microsphere. The smaller the waist size relative to the microsphere, the more it needs to be focused after the microsphere to reach the optimal $z_0$, which turns out to be beneficial for the intensity of the PNJ and therefore also for the NMS efficiency. Moreover, as the beam waist increases and surpasses the microsphere radius, part of the light gets wasted in the surroundings. Therefore, the increase in $w_0$ decreases the enhancement (**Figure 7**).

2. The microsphere radius $R$ has a low influence on the enhancement, which means it can still help to increase or decrease the performance of the NMS, but it is not critical for the optimization. Its limited effect is understandable, since its scope of influence is mainly limited to the PNJ generation, which is only of secondary importance to the contribution to the enhancement. Besides that, even its role in the PNJ generation is not critical when the incident beam position is optimized. [32] This low influence of microsphere size may seem like a surprise, since it is the size of the lens on which the light scattering is happening, but if the incident beam position is always optimal (like in our current numerical simulations), it will mitigate most of the differences related to $R$.
The microsphere refractive index $n_s$ has a significant influence on the enhancement. The dependence is almost a monotonous decrease except for a slight rise in enhancement for $n_s > 2.5$. It resembles the dependence from our previous research[33] about PNJ and high refractive index microspheres, but does not match it fully. There, if only optimal $z_0$ is tracked, the PNJ intensity drops from $n_s = 1.5$ but starts rising from $n_s = 1.8$. In the current situation, the enhancement decreases from (at least) $n_s = 1.33$, the rise starts only after $n_s > 2.5$. This indicates that $n_s$ not only influences the PNJ generation, but it also affects the plasmonic excitation, which in combination gives the resulting graph in **Figure 7**. The current data is insufficient to explain the observed behavior, but it is possible that a low refractive index material, such as $H_2O$ benefits the refraction of the light when it exits the microsphere into the cavity and onto the nanoelement more than a higher refractive index material such as $SiO_2$ and furthermore BTG.

3. Regarding the cavity parameters, the effect of its position $z_c$ on the enhancement can be explained by interpreting it together with the effect of the cavity radius $r_c$. The results show that the enhancement does not change significantly whether the cavity is heavily inside or outside the microsphere, or in between those positions. This is an interesting result, since the dependence on the cavity radius $r_c$ shows that the cavity is not irrelevant. The cavity needs to exist at least to make room for the tip, but not be comparable in size to the microsphere. If there is no cavity, the enhancement will be quenched substantially. This suggests that the primary role of the cavity is to house the nanoelement tip; as a light scattering aid, it has a limited effect. If the cavity is so large that it is comparable in size to the microsphere, it will heavily modify the geometry of the microsphere and consequently disturb its ability to generate the PNJ, which will lead to weaker plasmon excitation at the nanoelement and weaker overall enhancement. Moreover, to translate this to the explanation of the effect of $z_c$: the position of the cavity is not crucial since it houses the nanoelement tip in every position, while not modifying the microsphere geometry too much to hinder PNJ generation.

4. Lastly, the parameters of the nanoelement show widely different behaviors, from low to crucial influence on the enhancement. The low effect of the angle of the nanoelement from the vertical $\alpha$ on the enhancement arises from the distribution of the enhancement, which surrounds the nanoelement tip. Since the enhancement is present on all sides of the exposed tip, changing the angle of the nanoelement, which is effectively a rotation of the tip, will not make much difference in exciting the plasmons and generating the enhancement. Therefore, its effect will be limited.
On the other hand, the inner angle of the nanoelement $\theta$ significantly affects the enhancement. The results show that moderate values of $\theta$ are the best. The nature of this dependence can be explained by considering the character of plasmon excitation on metallic objects such as rods.[34] Here the lightning rod effect plays the main role. Similarly to its macroscopic analogue, where the electric field concentrates at the tip (or point) of a pointed metallic object, on the nanoscale the plasmonic excitement and electric field are massively concentrated at the tip. The effectiveness of this nanoscale lightning rod can be tailored by the geometry of the object, namely its inner angle and tip size.
The second geometry parameter of a pointed lightning rod, tip size, leads us to the effect of the tip radius $r_t$ on the enhancement, which showed by far the largest influence in the results. The smaller the $r_t$ is, the stronger the enhancement is, which is analogous to the occurrence that the pointier the lightning rod is, the stronger the concentration of the electric field will be.
In the end, we look at the nanoelement length parameter $h$. The reason why it suddenly quenches the enhancement when its value reaches the same order of magnitude as the microsphere size, could be due to the change of the role of the nanoelement. While it is small relative to the microsphere, it accomplishes its role of being a small metallic shape on whose tip the plasmons are being excited, while not interfering with the light scattering and PNJ generation from the microsphere. In that regime, the size of the nanoelement does not change the NMS performance much, which can be seen in the graph as the upper plateau of points (**Figure 8**). However, if the nanoelement becomes too large, it starts to occupy a significant part of the space inside the microsphere, which in turn interferes with the light refraction, PNJ generation, and consequently the plasmonic excitation.

**Proposed NMS control and usage in measurements**

The NMS could be controlled in the same way as simple microspheres were already controlled in the literature. By a laser trap[35] or by various mechanical holders, such as: AFM cantilever,[36,37] optical fiber,[15] micropipette,[38] holder for microsphere array,[39] microscope objective adapter[40] or even by direct adhesion of the microsphere to the microscope objective.[41] By incorporating such solutions, NMS could be used in mapping mode where every point in an image could be significantly enhanced in both intensity and resolution in methods such as Raman, infrared or fluorescence spectroscopy. The envisaged device and usage could potentially be as useful and impactful as TERS or potentially even more, but with the great benefit of being cheaper to produce and use. In TERS, one of the main problems is the difficulty regarding the positioning of the incident laser beam on the apex of the tip and the collection of the scattered signal[1]. In contrast to TERS, NMS would not need another system such as SPM for its functioning because the plasmonic contribution is contained inside the cavity of the microsphere, and also there is no need for fine control and the feedback loop like in the case of SPM tip, since NMS would not be that fragile. Moreover, NMS would need only one incident light beam, which is coming perpendicularly through the NMS, onto the sample and back through the same way into the same microscope objective. This is different than many TERS systems, which require an additional microscope objective for the incident beam at an angle. Therefore, NMS could be fitted on almost any present Raman spectrometer without other investments in equipment. Since the nanoelement is securely attached and held inside the cavity of the microsphere, it would not be as fragile or subjected to such strong forces as the AFM tip, and therefore, one NMS could last much longer. This further reduces the cost and eases the use. Even if the cost of production of one NMS would be larger than one AFM tip, the NMS would be much less expendable which would in the end be cheaper.

The possible drawbacks of the NMS include a limit in the space on the sample where it can be used because of the large size of the microsphere compared to a TERS tip. Also, the microsphere could prevent the nanoelement from reaching deep crevices in the sample. Both of these could be mitigated by extending the nanoelement or positioning it in a way that it protrudes outside of the outer edge od the microsphere. In that case, a system for fine control and the feedback loop would be needed to track the distance from the surface.

The NMS could be used in many micro-Raman applications in single-point acquisition and mapping. This includes for example material science investigations of semiconductor nanowires, carbon nanotubes and graphene, chemical identification of various compounds in mixture, examination of minerals, polymers, biological samples and other.

**Conclusions**

A new optical device called nano-engineered microsphere (NMS) has been presented, whose purpose is to acquire signal and resolution enhancement in methods such as Raman, infrared or fluorescence spectroscopy. NMS consists of a dielectric microsphere, containing a cavity on the shadow side, and a plasmonic nanoelement inside the cavity. The incident light is enhanced and focused multiple times: a PNJ is generated at the shadow side of the microsphere, which excites plasmons on the nanoelement, which enhance the signal and the resolution of mapping at the sample surface. At the collection path, the microsphere acts as an antenna, further enhancing the signal.

The parameters of the NMS, which are: the parameters of the incident beam (wavelength $\lambda$, beam waist radius $w_0$ and beam waist position $z_0$), the microsphere (radius $R$ and refractive index $n_s$), the cavity (radius $r_c$ and distance from the microsphere center $z_c$) and the nanoelement (angle relative to the vertical axis $\alpha$, inner angle $\theta$, length $h$, complex refractive index $n_n$, and radius of the curvature of the tip $r_t$), have been optimized by numerical simulations to obtain the highest enhancement of the electric field. The parameters have shown a wide level of influence on the enhancement, from low to critical. Microsphere radius, nanoelement angle from the vertical and cavity position have a low effect. The dependency of the enhancement on the wavelength and nanoelement angle are curves with a maximum, being the highest when the wavelength is around 650 nm and the nanoelement angle is of moderate value around 60º. On the other hand, the dependency curves for the incident beam waist radius and radius of the nanoelement tip fall exponentially, meaning that by increasing those parameters, the enhancement sharply drops, especially in the case of the tip radius, which has the biggest influence on the enhancement. In the case of the cavity radius and the length of the nanoelement, there is a certain narrow range of values to have a substantial enhancement. Otherwise, the enhancement is highly quenched. The dependencies on the microsphere and nanoelement material are more complicated. The best material for the microsphere is $SiO_2$, while the best material for the nanoelement is silver.

The proposed NMS device could be mounted and controlled by some of the currently available methods from the literature. In this configuration, NMS could be used in imaging measurements competing with TERS method but with the benefit of a much lower price and potentially stronger signal. Along with the presentation of a novel device and its optimization, this research lays the groundwork and writes the recipe for the experimental realization of the device. Moreover, it also opens up a new pathway in the field of combined enhancement by plasmonics and PNJ, and in the field of Raman scattering enhancement methods. If experimentally realized in the future, NMS could become the leading method of enhancement of the signal intensity and the spatial resolution in Raman and infrared spectroscopy.

**Acknowledgements**

The authors acknowledge financial support from the European Regional Development Fund for the project 'Materials for clean energy, advanced sensors and quantum technologies’ (Grant No. PK.1.1.10.0002).

The authors acknowledge financial support from the Croatian Science Foundation (HRZZ) through European Union National Recovery and Resilience Plan (NPOO – framework C3.2. R2-I1) for mobility programe for research stay grant MOBODL-2023-08-2860.

The work was supported by the Ministry of Innovation and Technology of Hungary from the National Research, Development and Innovation Fund (NKFIH) via TKP2021-NVA-04, and K146733 grants.

**Data availability**

The data that support the findings of this study are available from the corresponding author upon reasonable request.

**Conflict of interest**
The authors declare no conflict of interest.